\documentclass[aps,prb,reprint,longbibliography]{revtex4-2}
\usepackage{graphicx} 
\usepackage{dcolumn}
\usepackage{amsmath}
\usepackage{amssymb}
\usepackage{bm}

\begin{document}
\title{Dzyaloshinskii-Moriya interaction torques and domain wall dynamics in van der Waals heterostructures}
\author{Jun Chen}
\altaffiliation{These authors contributed equally to this work}
\author{Churen Gui}
\altaffiliation{These authors contributed equally to this work}
\author{Shuai Dong}
\email{Email: sdong@seu.edu.cn}
\affiliation{Key Laboratory of Quantum Materials and Devices of Ministry of Education, School of Physics, Southeast University, Nanjing 211189, China}
\date{\today}

\begin{abstract}
Since the discovery of two-dimensional ferroelectric and ferromagnetic materials, the van der Waals (vdW) heterostructures constructed by ferroelectric and ferromagnetic monolayers have soon become the ideal platforms to achieve converse magnetoelectric functions at the nanoscale, namely to use electric field to control magnetization. In this Letter, by employing density functional theory calculations and dynamic simulations of atomic spin model, we study the key role of interfacial Dzyaloshinshii-Moriya interaction (DMI) in CrI$_3$-In$_2$Se$_3$ vdW heterostructures. Our work demonstrates feasible DMI torques pumped by ferroelectric switching, which can drive current-free and low-energy consumption domain wall motion. Moreover, such interfacial DMI can also significantly enlarge the Walker field in magnetic field-driven domain wall technique.  
\end{abstract}

\maketitle

Originated from spin-orbit coupling (SOC), Dzyaloshinshii-Moriya interaction (DMI) plays a vital role in modern spintronics. By coupling neighbor spins in the antisymmetric form $H_D=\textbf{D}\cdot (\textbf{S}_i\times\textbf{S}_j)$ ~\cite{moriya1960anisotropic}, DMI can lead to many noncollinear spin textures, e.g. canting antiferromagnetism, chiral domain walls, as well as skyrmions ~\cite{bak1980theory,heide2008dzyaloshinskii,pappas2009chiral,muhlbauer2009skyrmion,do2009skyrmions}. It is well known that noncollinear spin textures usually produce nontrivial spin-electron scattering for transportation, which can contribute to topological (quantum) Hall effects and motion of quasi-particles~\cite{ohgushi2000spin,taguchi_spin_2001,onoda_spin_2003,parkin_magnetic_2008,iwasaki2013universal,barker2016static}, facilitating functional manipulations at the nanoscale. Because the presence of DMI relies on the lack of space-inversion symmetry, DMI widely exists in those non-centrosymmetric magnetic materials.  In particular, helimagnets such as Mn$_x$Fe$_{1-x}$Si, Fe$_x$Co$_{1-x}$Si, and FeGe, in which the DMI causes  spiral orders and/or skyrmions, have attracted lots of attentions~\cite{Pfleiderer_2010,wilhelm_precursor_2011,yu2011near,viennois_observation_2015,mcgrouther_internal_2016}. Besides, the asymmetric interfaces in artificial heterostructures or superlattices, can also produce the interfacial DMI which leads to various emergent phenomena~\cite{bode2007chiral,heide2008dzyaloshinskii, boulle_room-temperature_2016,woo_observation_2016,fert2017magnetic,yang_significant_2018,soumyanarayanan_emergent_2016}.

Reversibility of DMI vector provides a promising route to control the chirality of magnetism. Multiferroics allows the coexistence of more than one ferroic order within a single phase material, which provide a platform to control DMI vector via electric field or mechanical methods~\cite{dong_multiferroic_2015,cheong_multiferroics_2007,dong_magnetoelectricity_2019}. For some specific multiferroic systems, e.g. BiFeO$_3$, due to the locking between oxygen octahedral distortions and dipole moments, the DMI vector can be reversed by ferroelectric switching, leading to the fipping of canting moment~\cite{heron_electric-field-induced_2011,heron_deterministic_2014}. By reversing the DMI vector, the topological spin textures or solitons can be manipulated~\cite{srivastava2018large,liang2020electrically,xu2020electric,ba2021electric}. Following this mechanism, Yu {\it et al} demonstrated the chirality-dependent skyrmion-skyrmion interactions~\cite{nwac021}.  

The magnetic chirality also widely exists in domain walls, and thus DMI also plays an essential role in domain wall dynamics~\cite{thiaville_dynamics_2012,yoshimura_soliton-like_2016,vandermeulen_effect_2016,garcia_magnetic_2021,brandao_understanding_2017,dai2023electric}. For example, the reversal of DMI vector can produce a dissipative transverse motion during the chirality switching, as demonstrated in perovskite multiferroics ~\cite{chen2021manipulation,li_ultralow_2023}. Comparing with the scenario based on current-dependent torques~\cite{slonczewski1996current,berger1996emission,vzelezny2014relativistic}, such a pure electric field control has the advantage for its lower energy consumption, which has been coined as a new class of ``DMI torques''~\cite{yu2023voltage}. 

\begin{figure}[t]
    \centering
    \includegraphics[width=0.48\textwidth]{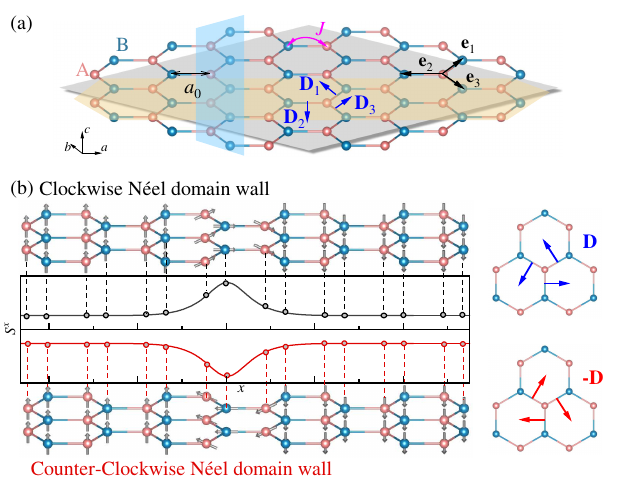}
    \caption{(a) Schematic honeycomb lattice of magnetic layer stacking on a substrate, where the exchange interaction $J$ and interfacial DMI vector $\textbf{D}_\mu$  ($\mu=1, 2, 3$) are located at each nearest neighboring bond between A site and B site. $\textbf{e}_\mu$: normalized axes. (b) The N\'eel-type domain walls with clockwise or counter-clockwise chiralities, plotted as one-dimensional coordinate functions of spin component $S^x(x)$. Such chirality can be determined by the DMI vector correspondingly.}
    \label{hc_dw}
\end{figure}

\begin{figure*}[t]
    \centering
    \includegraphics[width=1\textwidth]{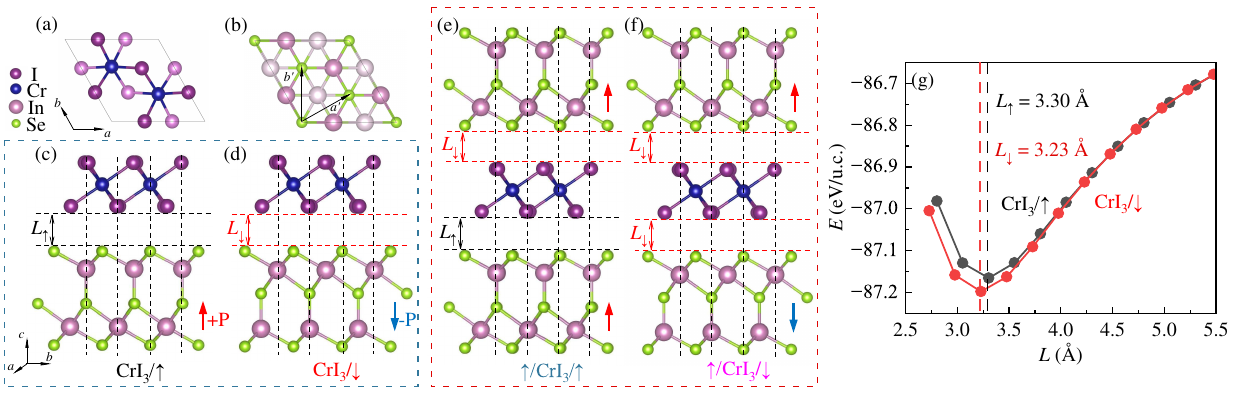}
    \caption{Structures of CrI$_3$-In$_2$Se$_3$ heterostructures. (a-b) Top view of CrI$_3$ monolayer (unit cell) and $\sqrt{3}\times\sqrt{3}$ supercell of $\alpha$-In$_2$Se$_3$. (c-f) Side views of heterostructures. (c) CrI$_3$/$\uparrow$, (d) CrI$_3$/$\downarrow$, (e) $\uparrow$/CrI$_3$/$\uparrow$ and (f) $\uparrow$/CrI$_3$/$\downarrow$, in which $\uparrow$ or $\downarrow$ denotes the polarization direction in In$_2$Se$_3$ layer. (g) The total energy as a function of interlayer distance $L$ for CrI$_3$/$\uparrow$ and CrI$_3$/$\downarrow$, where $L_\uparrow$ and $L_\downarrow$ represent their optimal interlayer distances.}
    \label{CrI/InSe}
\end{figure*}

Van der Waals (vdW) heterostructures are ideal platforms to design electronic devices with controllable DMI, benefiting from their high-usage interfaces and easy control of stacking modes. As a widely studied system, $\alpha$-In$_2$Se$_3$ monolayer was theoretically predicted to be ferroelectric ~\cite{ding_prediction_2017}, which was then fabricated and confirmed in experiments~\cite{zhou_out--plane_2017}. After that, In$_2$Se$_3$-based heterostructures have been theoretically studied and designed for nonvolatile electric field manipulations of various physical properties, including magnetism~\cite{cheng_nonvolatile_2021,wang_switchable_2022}, topological states or spin textures~\cite{zhang_heterobilayer_2021,huang_ferroelectric_2022,shen2023manipulation,wang_switching_2024}, band alignments and charge transfer~\cite{hu_ferroelectric_2022,zhang_ferroelectric_2024}. 
Although domain walls have been attracting more and more attentions in two-dimensional systems~\cite{wahab_quantum_2021,abdul-wahab_domain_2021,alliati_relativistic_2022}, the DMI control of domain wall dynamics remains less touched.

In this Letter, the DMI torques and domain wall dynamics of CrI$_3$-In$_2$Se$_3$ vdW heterostructures are studied. Our first-principles calculations indicate the interfacial DMI in CrI$_3$ monolayer can be induced in these heterostructures and the full reversal of DMI vector can be achieved in the In$_2$Se$_3$/CrI$_3$/In$_2$Se$_3$ sandwich-like heterostructure, which can generate DMI torques to drive the motion of domain walls. Furthermore, the existing of interfacial DMI can enlarge the critical field of Walker breakdown in magnetic field-driven domain wall motion. Such magnetoelectric vdW heterostructures shed lights on the low energy-consuming domain wall nanoelectronic applications.

CrI$_3$ monolayer is a well-recognized two-dimensional ferromagnetic sheet with a magnetic easy axis perpendicular to its plane ~\cite{huang_layer-dependent_2017}. The effective spin Hamiltonian for CrI$_3$'s honeycomb lattice can be written as:
\begin{equation}
    H=-J\sum_{<i,j>}{\textbf{S}_i\cdot\textbf{S}_j}-\sum_{<i,j>}\textbf{D}_{ij}\cdot(\textbf{S}_i\times\textbf{S}_{j})-A\sum_i({S_i^z})^2,
    \label{Hm_d}
\end{equation}
where $\textbf{S}$ is the normalized spin vector, $i$/$j$ are site indexs, and $<>$ denotes nearest-neighbors. As depicted in Fig.~\ref{hc_dw}(a). The first term is the ferromagnetic exchange ($J>0$), and the second term describes the interfacial DMI with the coefficient vector $\textbf{D}$. Due to the inversion symmetry, $\textbf{D}=0$  for isolated CrI$_3$ monolayer. However, the symmetry can be broken from the original $D_{3d}$ to $C_{3v}$ for CrI$_3$ monolayer on substrate, producing nonzero DMI vectors satisfy $\textbf{D}_\mu=\textbf{e}_z\times\textbf{e}_\mu$  ($\mu=1, 2, 3$) as depicted in Fig.~\ref{hc_dw}(a). The last term is the magnetocrystalline anisotropy with the easy axis along the $z$-axis ($A>0$). 

Due to the presence of DMI, the magnetic domain walls should belong to the N\'eel-type with specific chirality (i.e. clockwise (CW) or counter-clockwise (CCW)) to reduce the energy cost from DMI, as sketched in Fig.~\ref{hc_dw}(b).  In such a honeycomb lattice, although domain walls are possibly propagating along the zig-zag or armchair directions, i.e. the shadow stripes in Fig.~\ref{hc_dw}(a), they can be proved being equivalent in the magnetic dynamics~\cite{SM}. Then in the continuous limit, the model Hamiltonian for N\'eel-type domain walls propagating along the $x$-axis can be expressed as:
\begin{equation}
    H=\int{[\frac{J}{2}(\nabla\textbf{S})^2-D\textbf{e}_y\cdot(\textbf{S}\times\nabla\textbf{S})-\frac{4A}{3}(S^z)^2]dx},
    \label{Hm_c}
\end{equation}
where $D$ is the amplitude of DMI vector. Therefore, the domain wall can be described as $S^x=\text{sech}(x/\Delta)$ with the characteristic width $\Delta=\sqrt{3J/8A}$ ~\cite{Landau_1935}. It's worthy to note that the effective width of domain walls in the honeycomb lattice should be $\pi\Delta a_0$ ($a_0$ is the Cr-Cr distance as defined in Fig.~\ref{hc_dw}(a)).

\begin{figure}[t]
    \centering
    \includegraphics[width=0.48\textwidth]{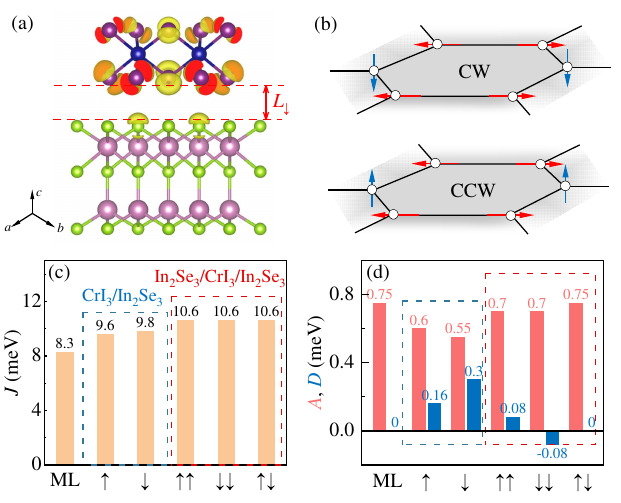}
    \caption{(a) Charge redistribution between the upper I and lower I surfaces in CrI$_3$/$\downarrow$. The charge density is plotted near the Fermi level with range of [$-0.3$, $0$] eV with isosurface level as $0.005$ $e$/\AA$^3$. (b) Noncollinear Cr-spin orders in honeycomb lattice of CrI$_3$ layer, where out-of-plane (blue arrows) and in-plane (red arrows) spins are in arrangements with different chiralities CW or CCW. (c, d) The exchange coefficient $J$, magnetocrystalline anisotropy $A$, and strength of DMI vector $\textbf{D}$ in CrI$_3$ monolayer (ML) and CrI$_3$-In$_2$Se$_3$ heterostructures, where $\uparrow$ or $\downarrow$ denotes the polarization of In$_2$Se$_3$ layer.}
    \label{DMI}
\end{figure}

To obtain the effective coefficients from real heterostructures, the first-principles calculations based on density functional theory (DFT) are performed, and technical details of calculations can be found in Supplementary Materials (SM)~\cite{SM}. Both CrI$_3$ and In$_2$Se$_3$ monolayers are hexagonal lattices, as shown in Fig.~\ref{CrI/InSe}(a-b), and their lattice constants are well-matched by following the (1$\times$1)/($\sqrt{3}\times\sqrt{3}$) stacking mode with less than $0.5\%$ mismatch. The CrI$_3$/In$_2$Se$_3$ bilayer and In$_2$Se$_3$/CrI$_3$/In$_2$Se$_3$ trilayer are considered. Here, the energetically most favorable stacking modes are displayed in Fig.~\ref{CrI/InSe}(c-f), while other stacking modes are discussed in SM~\cite{SM}. The CrI$_3$/In$_2$Se$_3$ bilayer is polar despite the polarization direction of In$_2$Se$_3$. The In$_2$Se$_3$/CrI$_3$/In$_2$Se$_3$ trilayer can be polar or nonpolar, depending on the combination of polarizations of two In$_2$Se$_3$ layers, as shown in Fig.~\ref{CrI/InSe}(e-f). For bilayer heterostructures CrI$_3$/$\uparrow$ and CrI$_3$/$\downarrow$ (here $\uparrow$ or $\downarrow$ denotes the polarization direction in In$_2$Se$_3$ layer), the energy is calculated as the function of interlayer distance $L$. As shown in Fig.~\ref{CrI/InSe}(g), their optimal distances $L_\uparrow$ and $L_\downarrow$ are only slightly different. Similar conclusion is also applicable to the trilayer heterostructures, in which the optimal $L$'s slightly depend on the polarization of In$_2$Se$_3$ layers.

As aforementioned, the DMI can be induced by the proximity effects at interfaces. Microscopically, DMI can originate from multiple factors, including the lattice distortion, orbital hybridization, electrostatic field, and charge transfer. Here, although the lattice distortions between Cr-I-Cr bonds are rather tiny after optimization, the charge density is significantly different between the upper I and lower I surfaces, due to the electrostatic field from In$_2$Se$_3$, as shown in Fig.~\ref{DMI}(a). To calculate the DMI vector, two noncollinear magnetic orders with identical exchange energy but opposite chiralities as depicted in Fig.~\ref{DMI}(b), are calculated with SOC. The energy difference between these two states is only contributed by the DMI according to $D=(E_\text{CCW}-E_\text{CW})/4$.  

All calculated coefficients including $J$, $D$, and $A$ are summarized in Fig.~\ref{DMI}(c-d). On one hand, comparing with monolayer CrI$_3$, the ferromagnetic exchange $J$ is strengthened by $15\%$-$28\%$ in these heterostructures, while the magnetocrystalline anisotropy $A$ is reduced by $0$-$26\%$. Thus the ferromagnetism of CrI$_3$ should remain robust. On the other hand, nonzero $\textbf{D}$'s appear only in those polar heterostructures. For bilayers, the amplitude of DMI's are estimated as $0.16$ meV and $0.3$ meV for CrI$_3$/$\uparrow$ and CrI$_3$/$\downarrow$ respectively. Interestingly, the sign of DMI does not change upon the ferroelectric switching, implying that the single interface itself contributes to $0.23$ meV while the reversable DMI contributes to $0.07$ meV. For trilayer, the DMI is completely reversed from $0.08$ meV in $\uparrow$/CrI$_3$/$\uparrow$ to $-0.08$ meV in $\downarrow$/CrI$_3$/$\downarrow$, while it is zero in $\uparrow$/CrI$_3$/$\downarrow$ or $\downarrow$/CrI$_3$/$\uparrow$ as expected from the inversion symmetry.

To study the spin dynamics of domain walls in the CrI$_3$ layer, we employed the atomic simulations by solving Landau-Lifshitz-Gilbert (LLG) equation~\cite{Landau_1935}: 
\begin{equation}
    \frac{\partial \textbf{S}}{\partial t}=\frac{\gamma}{\mu_s}\left(\textbf{S}\times\frac{\partial H}{\partial \textbf{S}}\right)+\alpha\left(\textbf{S}\times\frac{\partial \textbf{S}}{\partial t}\right),
\label{Eq2}
\end{equation}
where $\gamma=g\mu_{\rm B}/\hbar$ is the gyromagnetic ratio. $\mu_s=3$ $\mu_{\rm B}$ is the atomic magnetic moment for Cr$^{3+}$. The last term is for Gilbert damping with coefficient $\alpha$, which is typically determined by electron spin resonance in experiments. A stripy honeycomb lattice $N=600\times4$ is adopted, with periodic boundary conditions along the $y$-direction and open boundary conditions along the $x$-direction. The two $x$-ends are fixed as spin-up and spin-down. The fourth-order Runge-Kutta method is used to solve the LLG equation to obtain the time-dependent evolution of spins.  

\begin{figure}
    \centering
    \includegraphics[width=0.48\textwidth]{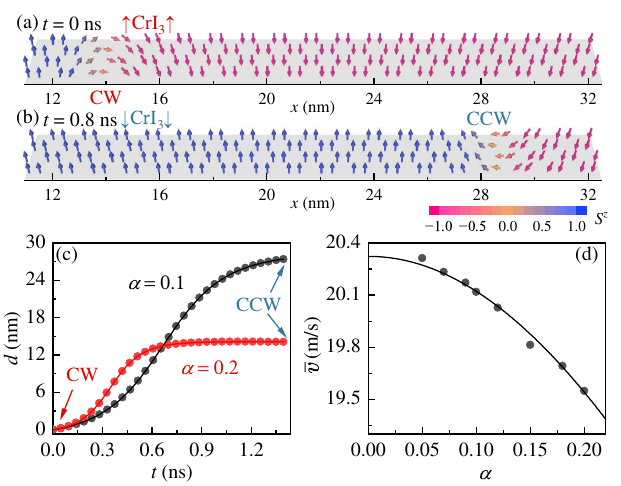}
    \caption{DMI torque-driven domain wall motion. (a-b) Snapshots of $\textbf{S}$ texture with an isolated domain wall, before and after the polarization flipping. The chirality of domain wall is reversed from CW (at $0$ ns) to CCW (at $0.8$ ns). (c) The time-dependent domain wall position $d$ with different damping coefficients. (d) The average velocities (solid circles) for the DMI torque driven-domain wall motion as a function of $\alpha$, which can be well fitted using the formula $v_m(\alpha^2+1)^{-1}$.}
    \label{DMIt}
\end{figure}

In the trilayer heterostructure, the DMI vector can be reversed by flipping the polarization, i.e. from $+0.08$ meV to $-0.08$ meV, which is anticipated to pump the DMI torques for spin dynamics~\cite{chen2021manipulation,yu2023voltage}. Here, we study an isolated domain wall at the honeycomb lattice, starting from a sharp boundary between the spin-up domain and spin-down domain. As shown in Fig.~\ref{DMIt}(a), due to the DMI ($D=0.08$ meV) in $\uparrow$/CrI$_3$/$\uparrow$, the final stable state is a CW N\'eel-type domain wall. After the polarization flipping to $\downarrow$/CrI$_3$/$\downarrow$, the DMI energy will produce an effective field acting on the domain wall, and thus pumps a DMI torque as:
\begin{equation}
{\bf \Gamma}_D=-\frac{2D\gamma}{\mu_s}(\textbf{S}\cdot\textbf{e}_y)\nabla\textbf{S},
\end{equation}
which will reverse the chirality of domain wall and induce a transverse motion as well. As shown in Fig.~\ref{DMIt}(b), with a large damping coefficient $\alpha=0.2$, the domain wall moves by $\sim15$ nm along the $x$-direction within $0.8$ ns, accompanying its chirality reversal to CCW. This dynamic process with a smaller damping coefficient $\alpha=0.1$ is also calculated for comparison. As shown in Fig.~\ref{DMIt}(c), the transverse motion distance $d$ reaches $\sim30$ nm within $\sim1.5$ ns. The motion of domain wall converges to stop gradually, thus it is not precise to define the motion time and the final distance   $d_s$. In fact, for the honeycomb lattice, the DMI torque-driven domain wall motion can be well described by:
\begin{equation}
    d=\frac{a_0\Delta}{\alpha}\text{arctan}[\text{sinh}(\frac{t}{Q})],
\end{equation}
where $Q=8\Delta(\alpha+\alpha^{-1})\mu_s/(3\pi\gamma D)$ denotes the characteristic time and the maximal moving distance is $d_s=a_0\Delta\pi/\alpha$. More details about the derivatives can be found in Supporting Information, and this kind of domain wall motion is similar to the "rolling-downhill"-type motion discovered in multiferroic perovskites~\cite{chen2021manipulation}. All the simulation results are in perfect agreement with the theory, in which the moving distance is proportional to $\alpha^{-1}$ and independent on the strength of DMI. Here, we can define the effective time for the motion as $\sim2\pi Q$, thus the average velocity can be calculated by $\bar{v}=3a_0\pi\gamma D/[16(\alpha^2+1)\mu_s]$. The average velocity $\bar{v}$ increases with decreasing $\alpha$ and its maximal limitation is $v_\text{m}=3a_0\pi\gamma D/16\mu_s$ when $\alpha\rightarrow0$, in consistent with the simulation results as shown in Fig.~\ref{DMIt}(d). For the In$_2$Se$_3$/CrI$_3$/In$_2$Se$_3$ heterostructure, the maximal average velocity is estimated as $\sim20$ m/s, which is close to the typical velocity of magnetic field-driven ferromagnetic domain walls below $50$ mT~\cite{schryer1974motion,jue_domain_2016,lin_universal_2016,yoshimura_soliton-like_2016}. 

\begin{figure}
    \centering
    \includegraphics[width=0.48\textwidth]{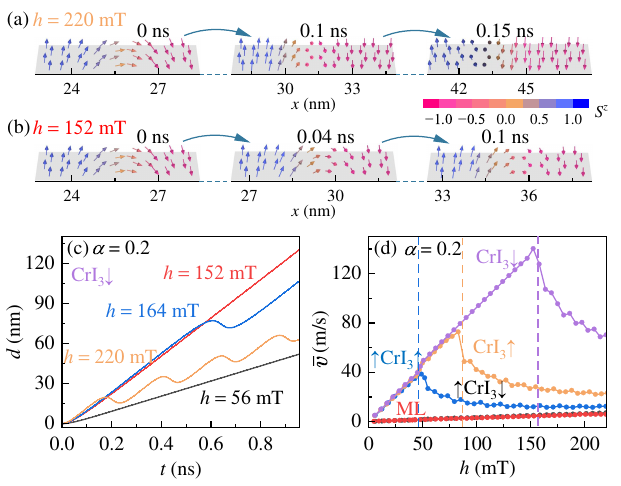}
    \caption{Magnetic field-driven domain wall motion with $\alpha=0.2$ .  (a-b) Snapshots of the $\textbf{S}$ textures in CrI$_3$/$\downarrow$ under different magnetic fields. Under the small field $152$ mT,  the dynamic process is mainly the steady motion, while it becomes a processional motion under $220$ mT. (c) The time-dependent domain wall positions under various magnetic fields. (d) Average velocity of domain wall motion as the functions of magnetic field for monolayer CrI$_3$ and CrI$_3$-In$_2$Se$_3$ heterostructures, where dash lines indicate the Walker fields calculated by Eq.~\ref{hw}.}
    \label{MHt}
\end{figure}

For completeness of domain wall motion, we also investigate the magnetic field-driven domain wall motion with different polarizations. Considering a magnetic field $h$ along the $z$-direction, the Zeeman energy term can be expressed as $H=-h\mu_s\sum_i{\textbf{e}_z\cdot\textbf{S}_i}$, which gives rise to a magnetic field torque as:
\begin{equation}
{\bf \Gamma}_h=\gamma h\textbf{e}_z\times\textbf{S}.
\end{equation}
The snapshots of domain wall dynamics in CrI$_3$/$\downarrow$ are displayed in Fig.~\ref{MHt}(a-b), where the magnetic field torque together with the Gilbert damping produce a dissipative torque to shift the domain wall and reduce the Zeeman energy. In this process, there is a critical field, called Walker field $h_\text{W}$~\cite{schryer1974motion}. Below $h_\text{W}$, the dynamic process is mainly the steady motion, while it becomes a processional motion above $h_\text{W}$ (Fig.~\ref{MHt}(d)). Here, the time-dependent domain wall motions under various magnetic fields are plotted in Fig.~\ref{MHt}(c-d), which indicates $h_\text{W}\sim152$ mT.

It is known that this $h_\text{W}$ orignates from the in-plane magnetic anisotropy. However, for CrI$_3$ monolayer itself, the fully occupied $t_{\rm 2g}$ orbitals at Cr ion together with the $D_{3d}$ point symmetry of honeycomb lattice, preserve the in-plane symmetry and thus the in-plane magnetic anisotropy is almost zero~\cite{jiang_controlling_2018,webster_strain-tunable_2018,kim_exploitable_2020}, as confirmed in our calculations. Therefore, the magnetic in-plane anisotropy comes from the interfacial DMI. With the coarse-graining approximation, the Walker field can be derived as:
\begin{equation}\label{hw}
    h_\text{W}=\frac{3\pi\alpha D}{8\Delta\mu_s},
\end{equation}
which agrees with above numerical value very well as shown in Fig.~\ref{MHt}(d).

Without this interfacial DMI, e.g. in CrI$_3$ monolayer, the domain wall motion can only be the processional motion, which leads to a velocity $v_p={a_0\alpha\gamma\Delta h}/{(1+\alpha^2)}$. In heterostructures, below $h_\text{W}$, the domain wall motion reaches a final steady velocity $v_s$ after a very short speeding duration ($\sim0.04$ ns). This $v_s={a_0\gamma\Delta h}/{\alpha}$ is much higher than the $v_\text{p}$. The numerical results of average velocity $\bar{v}$ are plotted in Fig.~\ref{MHt}(d) as a function of magnetic field for different heterostructures. Compared with those $D=0$ cases (CrI$_3$ monolayer and non-polar $\uparrow$/CrI$_3$/$\downarrow$), all polar heterostructures own larger $h_\text{W}$ and higher $\bar{v}$, as expected. In particular, below $h_\text{W}$, the velocity of domain wall can be up to $\sim140$ m/s for CrI$_3$/$\downarrow$, which is considerably large and similar to the gate-voltage effect in those magnetic film experiments~\cite{schellekens_electric-field_2012,lin_universal_2016}. 

In conclusion, by employing first-principles calculations and atomic spin dynamics simulations, we have studied the mechanism of interfacial DMI and domain wall dynamics in CrI$_3$-In$_2$Se$_3$ heterostructures. We have demonstrated the DMI torques can be pumped into In$_2$Se$_3$/CrI$_3$/In$_2$Se$_3$ trilayer heterostructures by polarization flipping, which contributes an efficient domain wall motion without either magnetic field or current. Furthermore, the existing and tunable DMI within heterostructures also plays an important role in improving the critical Walker field for magnetic field-driven domain wall motion, exhibiting the tunable and nonvolatile bias effect. In addition, our results are completely portable to other vdW heterostructures, which can be described by the effective spin model as well. Our results have performed rich manipulations of magnetic domain walls, which can be achieved in magnetoelectric vdW heterostructures.

\begin{acknowledgments}
This work was supported by the National Natural Science Foundation of China (Grant Nos. 12325401 \& 12274069), China Scholarship Council, Postgraduate Research \& Practice Innovation Program of Jiangsu Province (Grant No. KYCX21\_0079), and the Big Data Computing Center of Southeast University. 
\end{acknowledgments}

\bibliography{ref}
\end{document}